\pdfoutput=1
\documentclass[sigconf]{acmart}
\usepackage[english]{babel}
\usepackage{tabularx}
\usepackage{graphicx} 
\usepackage{algorithm} 
\usepackage{algpseudocode} 
\usepackage[table]{xcolor}
\usepackage{enumitem}
\usepackage{gensymb} 
\newboolean{showcomments}

\usepackage{booktabs}
\usepackage{array}
\usepackage{makecell}
\usepackage{subcaption}
\usepackage{multirow}
\usepackage[utf8]{inputenc}
\usepackage{newunicodechar}
\usepackage{fancyhdr}
\newcolumntype{Y}{>{\centering\arraybackslash}X}

\newunicodechar{ᵗ}{\textsuperscript{t}}
\newunicodechar{ʰ}{\textsuperscript{h}}
\AtBeginDocument{%
  \providecommand\BibTeX{{%
    Bib\TeX}}}
\def\BibTeX{{\rm B\kern-.05em{\sc i\kern-.025em b}\kern-.08em
    T\kern-.1667em\lower.7ex\hbox{E}\kern-.125emX}}

\copyrightyear{2026}
\acmYear{2026}
\setcopyright{cc}
\setcctype{by}
\acmConference[MM '26]{Proceedings of the 34th ACM International Conference on Multimedia}{November 10--14, 2026}{Rio de Janeiro, Brazil}
\acmBooktitle{Proceedings of the 34th ACM International Conference on Multimedia (MM '26), November 10--14, 2026, Rio de Janeiro, Brazil}
\acmDOI{10.1145/3767308.3835676}
\acmISBN{979-8-4007-2213-4/2026/11}

\begin{document}

\title[MD2G-Cast]{MD2G-Cast: Relay-Coordinated Multicast for Scalable Volumetric Streaming over MoQ}

\settopmatter{authorsperrow=4}

\author{Ruonan Chai}
\affiliation{%
  \institution{HKUST(GZ)}
  \department{Information Hub}
  \city{Guangzhou}
  \country{China}}
\email{rchai327@connect.hkust-gz.edu.cn}

\author{Yisu Wang}
\affiliation{%
  \institution{HKUST(GZ)}
  \department{Information Hub}
  \city{Guangzhou}
  \country{China}}
\email{ywang418@connect.hkust-gz.edu.cn}

\author{Zili Meng}
\affiliation{%
  \institution{HKUST}
  \department{Electronic and Computer Engineering}
  \city{Hong Kong}
  \country{China}}
\email{zilim@ust.hk}

\author{Dirk Kutscher}
\affiliation{%
  \institution{HKUST(GZ)}
  \department{Information Hub}
  \city{Guangzhou}
  \country{China}}
\email{dku@hkust-gz.edu.cn}

\renewcommand{\shortauthors}{Chai et al.}

\begin{abstract}
Volumetric streaming remains difficult to scale because receivers with overlapping fields of view are often served independently, causing repeated transmission of shared content. We present MD2G-Cast, a relay-coordinated multicast framework over Media over QUIC with an application-aware control layer for scalable multi-user volumetric delivery. MD2G-Cast jointly uses viewing overlap, receiver capability, and bandwidth conditions to form reusable multicast groups, share common Base content, and selectively admit Enhanced delivery. We formulate grouping and Enhanced admission as a sequential control problem, realize it with Proximal Policy Optimization (PPO), and train a compact relay model with teacher guidance for Enhanced admission. We implement MD2G-Cast with real MoQ processes and evaluate it with measured access traces and head-motion traces derived from a public 6DoF dataset for up to 100 users. At 20 and 100 users, MD2G-Cast keeps the receiver-side $P_{99}$ delivery interval below 40\,ms across all seven access profiles, while Rolling reaches the 500\,ms reporting cap in most cases. Across the evaluated user scales, MD2G-Cast achieves the highest or tied-highest mean system utility under homogeneous access and the highest mean utility under heterogeneous access, while reducing aggregate link load by about 27\% relative to Clustering at 100 users. A matched relay-control ablation separates the control structure from its optimizer, showing that random feasible actions reduce utility while deterministic control remains competitive with PPO. Together, the results support relay coordination and selective Enhanced admission, rather than a particular policy optimizer, as the central design contribution.
\end{abstract}



\maketitle

\section{Introduction}

\begin{figure}[t]
  \centering
  \includegraphics[width=\linewidth]{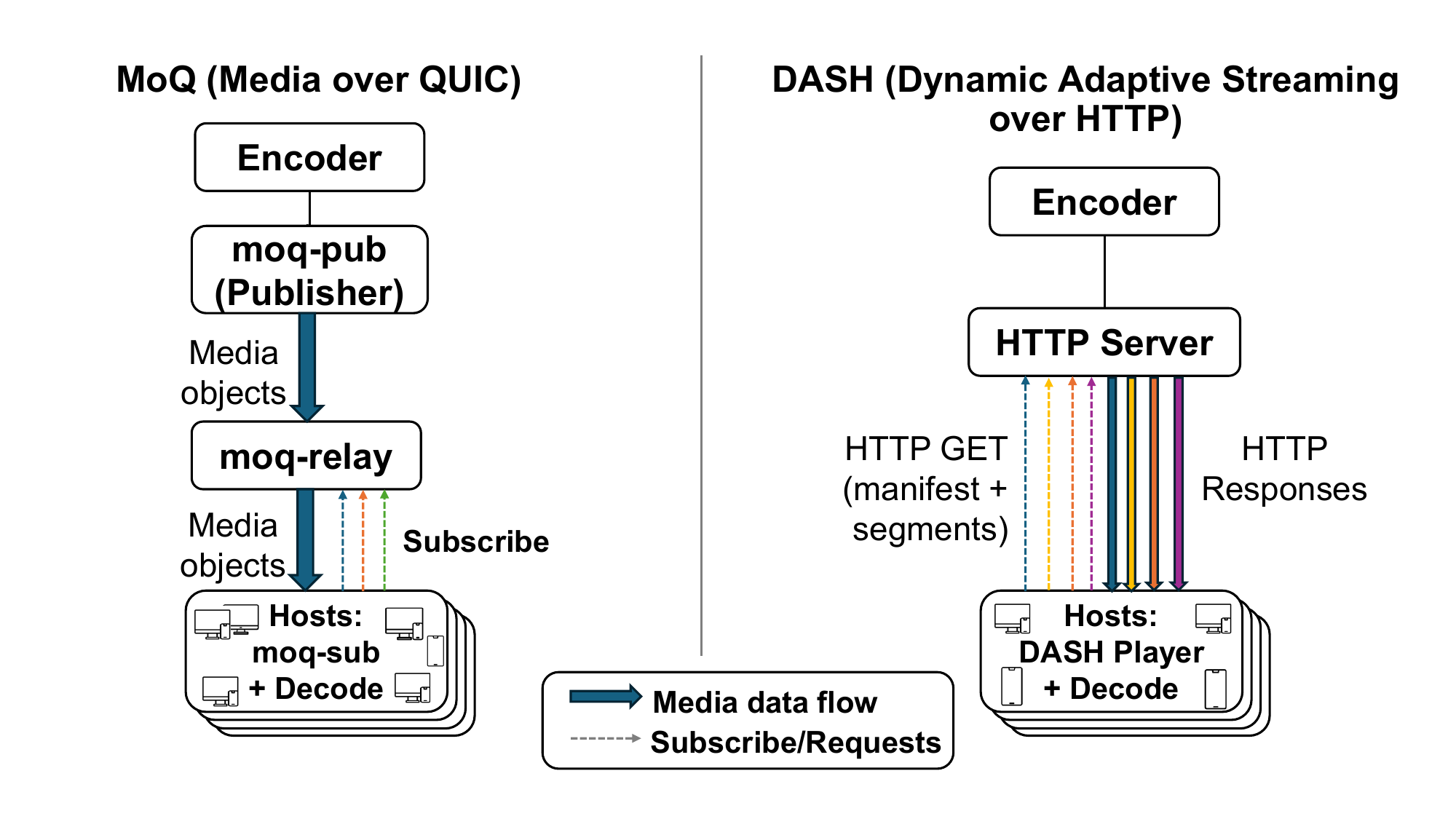}
  \caption{Pipeline contrast between DASH and MoQ for multi-user volumetric delivery.}
  \label{fig:moq-dash-pipeline}
  \Description{Comparison of independent DASH delivery and relay assisted MoQ delivery for multiple volumetric receivers.}
  \vspace{-1em}
\end{figure}

Immersive VR/AR applications increasingly use point cloud video to represent interactive 3D content. Unlike 2D or 360$^\circ$ video, point clouds encode scene geometry and support six-degrees-of-freedom, or 6DoF, navigation with continuously changing viewpoints and fields of view. Even after compression and viewport adaptation, high-fidelity volumetric streams can require hundreds of megabits per second~\cite{lee2020groot,li2022optimal,zhu2022semantic}. At scale, the challenge shifts from streaming to one user to serving many users with different views, devices, and links without wasting bandwidth on repeated content.

Most existing systems still follow a unicast mindset, optimizing each user independently through rate adaptation, tile-based encoding, or demand prediction~\cite{li2022optimal,lin2021deep,li2023toward,huang2023iscom,yeh2020mobile,zhu2022semantic,liang2024fumos,bian2024wireless,lin2022transformer}. These approaches improve per-user efficiency by reducing bitrate or anticipating future requests, but remain fundamentally user-centric. They therefore leave a multi-user bottleneck unresolved: substantial FoV overlap causes identical content to be repeatedly fetched while the network remains unaware of this redundancy. Figure~\ref{fig:moq-dash-pipeline} illustrates the resulting system-level difference. DASH relies on per-client HTTP requests and duplicates transfers even when users consume overlapping content, whereas MoQ allows shared media objects to be published upstream and replicated by relays toward subscribers with overlapping demand. This shift from request-driven to relay-coordinated delivery changes where redundancy can be eliminated in the network.

However, MoQ alone does not solve the multi-user control problem. It provides practical publish/subscribe and relay-assisted delivery over QUIC~\cite{10.1145/3587819.3593937,gurel2024media}, but leaves application decisions such as what content should be shared, how receivers should be grouped, and when additional quality should be admitted to the application. The remaining challenge is therefore to translate volumetric signals such as FoV overlap, receiver capability, and bandwidth headroom into coordinated relay decisions over shared Base content and optional refinement.

Multicast has long been used for 2D and 360$^\circ$ video by grouping receivers with similar views~\cite{perfecto2020taming,gao2024360,zhang2021joint,mahmoud2023survey}, but volumetric media introduces a different sharing problem. Volumetric delivery combines spatially sparse content, heterogeneous receiver capabilities, and layered representations in which broadly reusable content coexists with optional refinement. Recent systems have begun exploiting multi-user volumetric sharing~\cite{zhang2022m5,zhang2021hotnets,liu2024muv2,chai2025inds,zhang2022yuzu,liu2023cav3,ueno2023soft,lai2024socially}, yet many remain tied to particular grouping rules or tightly coupled adaptation mechanisms. This motivates a relay control abstraction that can exploit reuse while adapting grouping and refinement decisions as receiver and network conditions evolve.

MD2G-Cast addresses this gap with an application-aware control layer at MoQ relays. The control layer maps volumetric overlap and receiver state into a structured decision interface, dynamically forms multicast groups for shared Base delivery, and selectively admits Enhanced delivery when additional refinement is worthwhile. We formulate these coupled grouping and admission decisions as a Markov Decision Process and realize the sequential controller with Proximal Policy Optimization~\cite{schulman2017proximal}. Training uses real MoQ processes in a trace-driven Mininet environment, with teacher-guided KL matching applied to the compact model's Enhanced-admission branch~\cite{gou2021knowledge}. PPO is one realization of the control interface rather than the sole source of the system benefit.

\textbf{Our key contributions are as follows:}
\begin{itemize}[leftmargin=*, labelindent=0pt]
    \item We introduce a MoQ relay control layer that turns cross-user FoV overlap, receiver capability, and bandwidth headroom into coordinated grouping and Enhanced admission decisions. The transport and encoded media remain unchanged during relay control, separating the coordination problem from the underlying delivery substrate.

    \item We formulate grouping and Enhanced admission as a constrained sequential decision problem in which candidate construction enforces valid relay states before optimization, allowing learned and deterministic controllers to share the same relay interface.
    
    \item We implement a prototype with V-PCC content and real MoQ processes and evaluate it using same-substrate baselines, an independent-delivery reference, and a matched relay-control ablation that separates substrate, relay-control, and policy-realization effects.
\end{itemize}

Heuristic and Clustering share the MoQ substrate and measurement procedure with MD2G-Cast, whereas Rolling retains independent HTTP/DASH delivery and serves as a system-level independent-delivery reference rather than a transport-isolation baseline.

As shown in Section~\ref{sec:eval}, experiments with up to 100 concurrent users show that MD2G-Cast keeps the receiver-side $P_{99}$ delivery interval below 40\,ms across all seven access profiles at the 20- and 100-user loads reported in Fig.~\ref{fig:ttfb_latency}. MD2G-Cast also attains the highest mean system utility across the evaluated heterogeneous user scales and the highest or tied-highest utility under homogeneous access, while avoiding the largest aggregate link load incurred by Clustering.

\noindent\textbf{Artifact availability.}
The public artifact provides the implementation, Mininet orchestration scripts, trace manifests, and analysis code at \url{https://github.com/RuonanChai/MD2G_Cast}. Because the full media assets exceed the supplementary upload limit, the package contains shortened clips for end-to-end pipeline validation. These clips are not used to produce the reported measurements. All experimental results use the full 120-second Base and Enhanced workloads described in the supplementary material.
\section{System Design}
\label{sec:model}

MD2G-Cast is a relay-coordinated MoQ system for multi-user volumetric delivery. The VSS partitions each source point-cloud frame into complementary Base and Enhanced point sets and encodes them as independent V-PCC atlas streams. At runtime, relays use FoV overlap, receiver capability, and bandwidth headroom to coordinate multicast grouping and Enhanced admission without changing the encoded media. Table~\ref{tab:notation} summarizes the notation used below.

\begin{table}[t]
\centering
\caption{Key notation used in MD2G-Cast.}
\label{tab:notation}
\small
\setlength{\tabcolsep}{4pt}
\renewcommand{\arraystretch}{1.05}
\begin{tabularx}{\columnwidth}{@{}l X@{}}
\toprule
\textbf{Symbol} & \textbf{Description} \\
\midrule
$\mathcal{X}_f$ & Ordered source point set of a frame \\
$\mathcal{X}_b,\mathcal{X}_e$ & Complementary Base and Enhanced point sets \\
$\rho$ & Fixed Base sampling ratio used during media preparation \\
$\mathcal{I}_b$ & Deterministic index set selected for the Base partition \\
$\mathcal{U}_t$ & Active receiver set at control interval $t$ \\
$O(u_i,u_j)$ & Directional FoV overlap from user $u_i$ to user $u_j$ \\
$\mathcal{D}_{\mathrm{raw}}(u)$ & Aggregated raw device capability of user $u$ \\
$B_u$ & Measured bandwidth budget of user $u$ \\
$\mathcal{D}(u)$ & Normalized device capability of user $u$ \\
$\mathcal{B}(u)$ & Normalized bandwidth of user $u$ \\
$\mathcal{B}_{\mathrm{adq}}(u)$ & Relay-local bandwidth adequacy of user $u$ \\
$\bar{O}(u;C)$ & Mean overlap of user $u$ with candidate group $C$ \\
$\mathcal{G}(u;C)$ & Compatibility feature of user $u$ with candidate group $C$ \\
$\hat{\mathcal{G}}_t(u)$ & Relay-assigned grouping state of user $u$ at interval $t$ \\
$a_{e,t}(u)$ & Enhanced-admission indicator for user $u$ at interval $t$ \\
$\mathcal{S}_t$ & Aggregate relay state at control interval $t$ \\
$R_o,R_q,R_b$ & Grouping-efficiency, delivered-quality, and bandwidth-inefficiency terms \\
$U_i$ & Clipped system utility for sample or accounting window $i$ \\
$\lambda_d$ & Distillation coefficient for teacher-guided KL matching \\
\bottomrule
\end{tabularx}
\vspace{-1em}
\end{table}

\subsection{Architectural Overview}
\begin{figure*}[t]
  \centering
  \includegraphics[width=\textwidth]{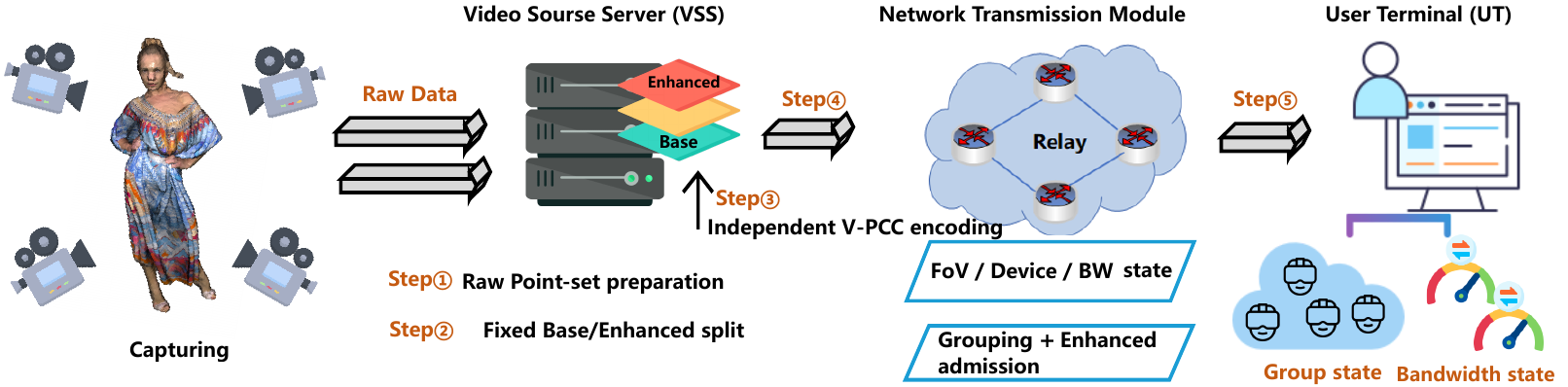}
  \caption{System architecture of MD2G-Cast.}
  \label{fig:system_architecture}
  \Description{MD2G-Cast architecture showing fixed source-side Base/Enhanced point-set partitioning, independent V-PCC encoding, and relay-side grouping and Enhanced admission driven by FoV, device, and bandwidth state.}
  \vspace{-1em}
\end{figure*}

Figure~\ref{fig:system_architecture} illustrates the system architecture. MD2G-Cast enables relay-side coordination without modifying the underlying transport protocol, allowing the decision logic to evolve independently of the transport stack. The system includes a Video Source Server (VSS), MoQ relays running the network transmission module, and User Terminals (UTs).

\subsection{Video Source Server (VSS)}

For each source frame, let $\mathcal{X}_f=\{x_1,\dots,x_n\}$ denote its ordered point set. For all reported experiments, the Base ratio is fixed at $\rho=0.6$ across access profiles and is not adapted online. With $m=\operatorname{round}(\rho n)$, a deterministic index-stride sampler selects a Base index set $\mathcal{I}_b$ containing $m$ points. The resulting point sets are
\begin{equation}
\mathcal{X}_b=\{x_i\mid i\in\mathcal{I}_b\},\qquad
\mathcal{X}_e=\mathcal{X}_f\setminus\mathcal{X}_b.
\label{eq:source_partition}
\end{equation}
By construction, $\mathcal{X}_b\cap\mathcal{X}_e=\varnothing$ and $\mathcal{X}_b\cup\mathcal{X}_e=\mathcal{X}_f$. The two point sets are encoded independently with V-PCC~\cite{8571288} into separate atlas streams. FoV overlap, receiver capability, and bandwidth do not affect this source-side partition; these signals are used later by the relay for grouping and Enhanced admission. Relays neither re-encode nor repartition the media during delivery. When both streams are available, the UT obtains the Base-plus-Enhanced reconstruction by stitching the decoded point subsets according to their original indices.

\subsection{Network Transmission Module}

MoQ relays scale delivery by grouping users with reusable demand and coordinating Base/Enhanced subscriptions. The directional FoV overlap from user $u_i$ to user $u_j$ is defined as
\begin{equation}
O(u_i,u_j)=\frac{|\mathcal{P}_{u_i}\cap\mathcal{P}_{u_j}|}{|\mathcal{P}_{u_i}|},
\end{equation}
where $\mathcal{P}_u$ denotes the set of patches visible in user $u$'s FoV. These FoV patches are used only to estimate cross-user viewing overlap and are distinct from the fixed Base/Enhanced point-index partition created at the VSS.

Relays further incorporate normalized device and bandwidth signals:
\begin{equation}
\mathcal{D}(u)=\frac{\mathcal{D}_{\text{raw}}(u)-D_{\min}}{D_{\max}-D_{\min}},\quad
\mathcal{B}(u)=\frac{B_u-B_{\min}}{B_{\max}-B_{\min}},
\end{equation}
where $\mathcal{D}_{\mathrm{raw}}(u)$ denotes the aggregated
device capability of user $u$, and $B_u$ denotes its measured
bandwidth budget.

For a feasible candidate group $C$ containing user $u$, we
define the directional overlap feature
\begin{equation}
\bar{O}(u;C)=
\begin{cases}
0, & |C|=1,\\[1mm]
\displaystyle
\frac{1}{|C|-1}
\sum_{v\in C\setminus\{u\}} O(u,v),
& |C|\ge 2.
\end{cases}
\label{eq:candidate_overlap}
\end{equation}
The corresponding compatibility feature is
\begin{equation}
\mathcal{G}(u;C)=
\gamma_1\bar{O}(u;C)
+\gamma_2\mathcal{D}(u)
+\gamma_3\mathcal{B}(u),
\qquad
\gamma_1+\gamma_2+\gamma_3=1.
\label{eq:group_score}
\end{equation}
For a fixed user, candidate dependence enters through
$\bar{O}(u;C)$, while $\mathcal{D}(u)$ and
$\mathcal{B}(u)$ describe receiver capability and current
bandwidth headroom. The relay supplies these features to the
policy for every feasible candidate. The policy, rather than
Eq.~\eqref{eq:group_score} alone, selects the final grouping
state. A singleton candidate remains a valid fallback.

\subsection{User Terminal (UT)}

The UT continuously decodes the Base atlas for baseline reconstruction. When Enhanced delivery is admitted and arrives before its playback deadline, the Enhanced atlas is decoded independently and its complementary point set is stitched with the decoded Base points using the original partition indices. Base and Enhanced therefore do not form a predictive decoding dependency, but the combined reconstruction uses the union of their complementary point subsets. Receiver quality is derived from the point-set configuration available at the UT and contributes to the runtime quality term defined in Section~\ref{sec:measurement}.

\section{Formulation and Solution}
\label{sec:formulation_and_solution}

\subsection{PPO Policy Realization}
\label{subsec:ppo_realization}

MD2G-Cast formulates relay grouping and Enhanced admission as a sequential control problem and uses PPO as one realization of this controller. PPO uses a clipped surrogate objective to moderate policy updates during training, while teacher-guided KL matching supervises the compact model's Enhanced-admission branch. The relay-visible state, feasible action structure, and coordinated grouping and admission interface define a common control abstraction for both learned and deterministic realizations. Section~\ref{subsec:ablation} compares learned and deterministic realizations under the same MoQ substrate.

\subsection{MDP Formulation}
Relay-local grouping and Enhanced admission are modeled as an MDP. At each control interval, the relay constructs a feasible candidate set from the current FoV relations and evaluates the compatibility features defined in Eq.~\eqref{eq:group_score}. The state also includes relay-local bandwidth adequacy, receiver capability, and the previous grouping assignment.

\textbf{State.}
\begin{equation}
\mathcal{S}_t(u)=
\left\{
O_t(u),
\mathcal{B}_{\mathrm{adq},t}(u),
\mathcal{D}(u),
\hat{\mathcal{G}}_{t-1}(u)
\right\},
\label{eq:mdp_state}
\end{equation}
where $O_t(u)$ summarizes the current overlap features over the feasible candidates, $\mathcal{B}_{\mathrm{adq},t}(u)$ denotes relay-local bandwidth adequacy, $\mathcal{D}(u)$ denotes normalized receiver capability, and $\hat{\mathcal{G}}_{t-1}(u)$ is the previous grouping state. 

Let $\mathcal{U}_t$ denote the active receiver set. The relay state is the collection $\mathcal{S}_t=\{\mathcal{S}_t(u):u\in\mathcal{U}_t\}$.

\textbf{Action.}
\begin{equation}
\mathcal{A}_t=\left\{\left(\hat{\mathcal{G}}_t(u),a_{e,t}(u)\right):u\in\mathcal{U}_t\right\},
\end{equation}
where $\hat{\mathcal{G}}_t(u)$ denotes the relay-assigned grouping state of user $u$, and $a_{e,t}(u)\in\{0,1\}$ denotes whether Enhanced delivery is admitted during the current control interval. Together, these two variables determine the service state delivered to the UT.

\textbf{Action realization.}
At $t=0$, grouping is initialized deterministically from FoV overlap, with singleton groups as a fallback. At each one-second control interval, the relay constructs valid relay-local candidate groups from the current overlap, bandwidth-adequacy, and device signals. The policy selects $\hat{\mathcal{G}}_t(u)$ and $a_{e,t}(u)$ only from this feasible set. Invalid assignments are rejected before subscription updates, and group migration is damped by hysteresis to avoid oscillation. The VSS generates fixed Base/Enhanced objects and their partition metadata before publication. Relays control grouping, subscription, and Enhanced admission but neither re-encode nor repartition content online.

\textbf{Feasibility before optimization.} Candidate construction enforces valid relay states before policy selection, separating correctness from optimization. PPO, deterministic rules, or another policy can therefore operate over the same relay state and action space without changing the encoded media or the MoQ transport path.

\textbf{Transition.}
\begin{equation}
\mathbb{T}(\mathcal{S}_{t+1}\mid \mathcal{S}_t,\mathcal{A}_t)
\end{equation}
captures relay-local state evolution under FoV and bandwidth dynamics.

\textbf{Implemented objective.}
For each runtime sample or accounting window $i$, the
implemented scalar objective is
\begin{equation}
U_i=
\operatorname{clip}_{[0,1]}
\left(
0.25R_{o,i}
+0.60R_{q,i}
-0.15R_{b,i}
\right),
\label{eq:system_utility}
\end{equation}
where $R_{o,i}$ is the normalized grouping efficiency term, $R_{q,i}$ is the bounded delivered quality term, and $R_{b,i}\ge0$ is the normalized bandwidth inefficiency term. Clipping is applied to each sample or accounting window before aggregation. Equation~\eqref{eq:system_utility} defines the system utility used throughout the evaluation and the matched relay control ablation.

\subsection{Distillation for Relay Execution}

We distill the Enhanced-admission branch of Student-128 from Teacher-512 using
\begin{equation}
L_{\mathrm{distill}}=\lambda_d\operatorname{KL}\!\left(\pi_t^e\Vert\pi_s^e\right),
\label{eq:distillation}
\end{equation}
where $\pi_t^e$ and $\pi_s^e$ are the teacher and student Bernoulli admission distributions induced by their sigmoid outputs, and $\lambda_d$ is the distillation coefficient. The critic is trained separately by standard PPO value regression against returns.

\section{Implementation}
\label{sec:implementation}

We implement MD2G-Cast on the open-source moq-rs codebase\footnote{\url{https://github.com/moq-dev/moq}}. The system design is described in Section~\ref{sec:model}.

\subsection{Relay-Coordinated Delivery}

The relay-side controller periodically observes FoV overlap, device capability, and bandwidth. It groups users with similar demand for shared Base delivery and selectively admits Enhanced delivery when resources permit. Grouping is updated at fixed control intervals, with hysteresis providing stability while overlapping subscriptions share upstream traffic.

\subsection{Client Behavior and Measurement}
\label{sec:measurement}

Each UT continuously receives and decodes the Base atlas stream. When Enhanced delivery is admitted and arrives before its playback deadline, the UT also decodes the Enhanced atlas stream and stitches its complementary points with the Base reconstruction using the original partition indices. The streams remain independent transport objects throughout delivery. Incoming useful media objects are timestamped on a receiver-side measurement path that is separated from playback processing.

The receiver-side delivery metric is an exponential moving average of inter-arrival intervals between useful payloads. The same fallback rules and the same $[1,500]$\,ms bounds are applied to every strategy. This metric represents sustained delivery continuity and tail inter-arrival behavior. It is not network RTT and not end-to-end capture-to-render latency.

For each user $u$ and accounting window $i$, the bounded
quality term is
\begin{equation}
R_{q,i}(u)=\operatorname{clip}_{[0,1]}\left(Q_{\mathrm{del},i}(u)-0.2P_{d,i}(u)-0.3P_{s,i}(u)\right),
\label{eq:quality_term}
\end{equation}
where $Q_{\mathrm{del},i}(u)$ is the normalized quality of the point-set configuration available at the receiver. $P_{d,i}(u)$ and $P_{s,i}(u)$ are the normalized delivery-interval and cumulative-stall penalties, respectively.

For reported utility, we discard the first 30\,s as warm-up, apply Eq.~\eqref{eq:system_utility} to each sample or window, and then compute the expected-client mean. An expected client without a valid served sample contributes zero. The same normalization, clipping, warm-up, and aggregation procedure is used for every strategy.

\section{Experimental Evaluation}
\label{sec:eval}
\begin{figure}[t]
  \centering
  \includegraphics[width=\linewidth]{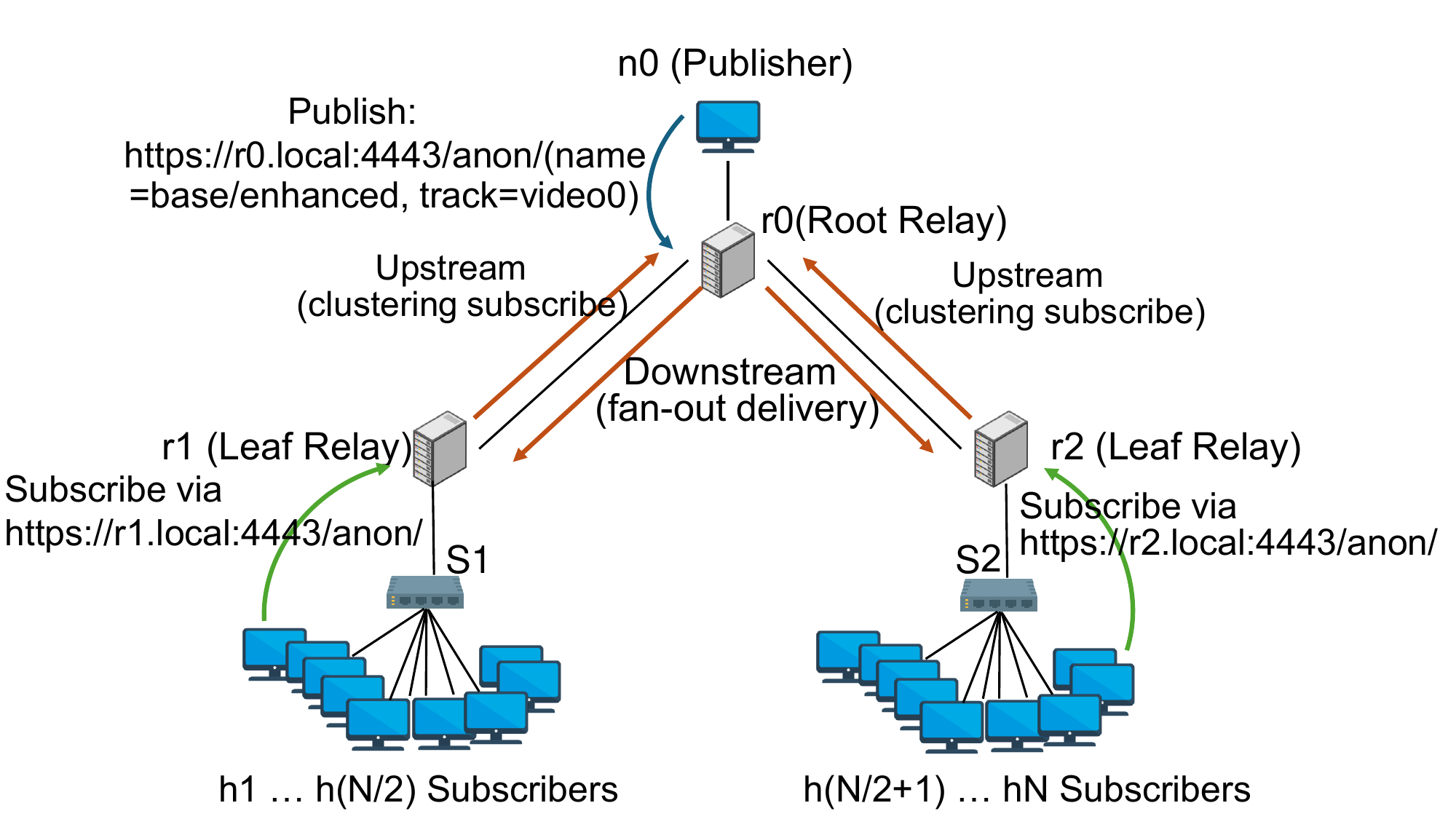}
  \vspace{0.6em}
  \includegraphics[width=\linewidth]{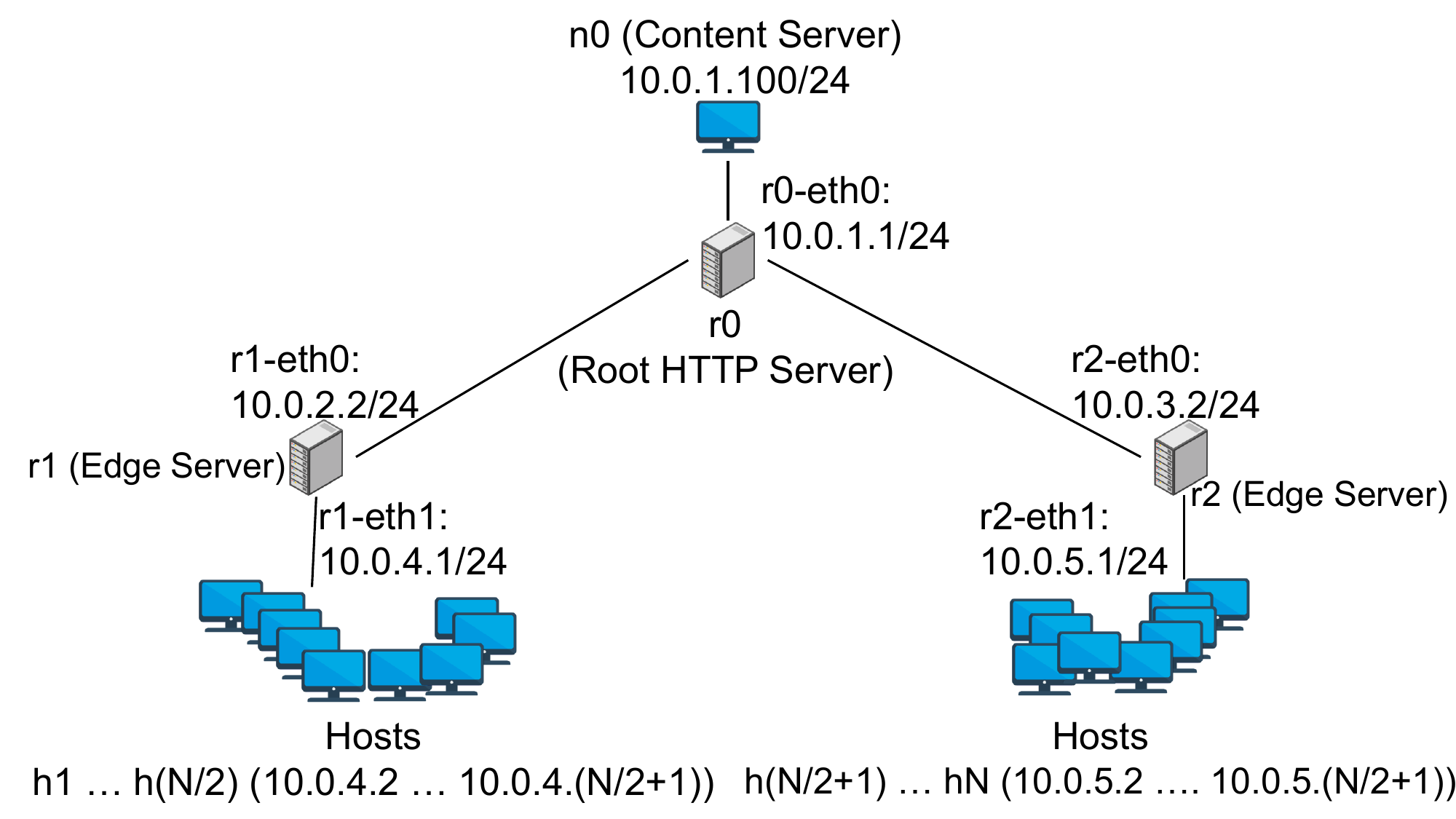}
  \caption{
  Experimental topology used for baseline comparison.
  Top: MoQ baseline topology. Bottom: DASH baseline topology.
  }
  \Description{Experimental MoQ and HTTP DASH topologies used for the controlled and independent delivery comparisons.}
  \label{fig:topologies}
\vspace{-1em}
\end{figure}

Our experiments are designed to validate three main claims:
\begin{enumerate}[leftmargin=*, labelindent=0pt]
    \item Relay-side policy inference must remain comfortably within the one-second control interval used by the system.
    \item Receiving complementary Enhanced content in addition to Base delivery must provide measurable reconstruction-quality improvement.
    \item Under trace-driven viewing dynamics and heterogeneous access conditions, MD2G-Cast should stabilize object delivery and provide a favorable system-utility trade-off without incurring excessive aggregate link load.
\end{enumerate}

We additionally use a matched relay-control ablation to separate the effects of dynamic grouping and action selection from the common MoQ delivery substrate.

\textbf{Testbed and topology.}
Figure~\ref{fig:topologies} shows our trace-driven Mininet testbed on a 32-vCPU server with 160\,GB RAM, where the MoQ publisher, relays, and subscribers run as independent processes. The topology contains one publisher, one root relay, two leaf relays, and up to 100 clients distributed across two access subnets. All strategies use the same host--relay graph, workload timing, content corpus, access traces, and measurement procedure.

\textbf{Bandwidth traces.}
We replay the real application-layer throughput traces, using 4G/5G traces from public Kaggle datasets,\footnote{\url{https://www.kaggle.com/datasets/kimdaegyeom/5g-traffic-datasets}; \url{https://www.kaggle.com/datasets/aeryss/lte-dataset}}
Wi-Fi traces from Zenodo~\cite{9318407},\footnote{\url{https://zenodo.org/records/6884095}}
and fiber traces from the FCC Measuring Broadband America program.\footnote{\url{https://www.fcc.gov/reports-research/reports/measuring-broadband-america/raw-data-measuring-broadband-america-eighth}}
We filter inconsistent samples, remove zero-throughput artifacts from the traces, and align them to the relay decision interval.

\textbf{Device capability.}
We model device heterogeneity from official specifications of representative HMDs,\footnote{\url{https://www.meta.com/}; \url{https://www.apple.com/apple-vision-pro/}} including CPU/GPU, memory, refresh rate, and display resolution; these features are normalized and used by the relay controller for grouping and Enhanced admission.

\textbf{Head movement.}
We derive head-motion traces from the CWI DIS 6DoF navigation dataset, selecting the H3 \textit{Red and Black} sequence from 8i~\cite{subramanyam2020user}. To isolate FoV overlap under controlled conditions, client positions are aligned to a common reference while the original orientation trajectories are preserved. The evaluation therefore retains realistic viewing-direction dynamics without claiming full translational 6DoF motion.

\textbf{Baselines and controlled variables.}
MD2G-Cast, Heuristic, and Clustering use the same MoQ publisher, relay, and subscriber processes, the same Base/Enhanced tracks, the same physical topology, and the same trace inputs and metric definitions. These methods form the controlled comparison for relay-side grouping and admission logic.

Rolling uses the same host--relay graph, client population, content timing, access traces, run duration, and measurement methodology, but retains an HTTP/DASH independent-delivery model with per-client fetching. Rolling therefore provides a system-level independent-delivery reference. It does not isolate the transport effect between HTTP/DASH and MoQ.

\begin{table}[t]
\centering
\caption{Inference profile of Teacher-512 and Student-128.}
\label{tab:model_profile}
\small
\setlength{\tabcolsep}{2pt}
\renewcommand{\arraystretch}{1.08}
\begin{tabular*}{\columnwidth}{@{\extracolsep{\fill}}lcccc@{}}
\toprule
\makecell[c]{\textbf{Model}\\\phantom{\textbf{(ms)}}} &
\makecell[c]{\textbf{\#Params}\\\phantom{\textbf{(ms)}}} &
\makecell[c]{\textbf{Latency}\\\textbf{(ms)}} &
\makecell[c]{\textbf{Inference rate}\\\textbf{(/s)}} &
\makecell[c]{\textbf{Memory}\\\textbf{(MB)}} \\
\midrule
Teacher-512 & 468{,}580 & 0.083--0.084 & $\sim$12k & 1.79 \\
Student-128 & 68{,}068 & 0.039--0.041 & $\sim$25k & 0.26 \\
\bottomrule
\end{tabular*}
\end{table}

\begin{figure}[t]
\centering
\subfloat[Base only]{%
    \includegraphics[width=2.2cm,height=4.2cm]{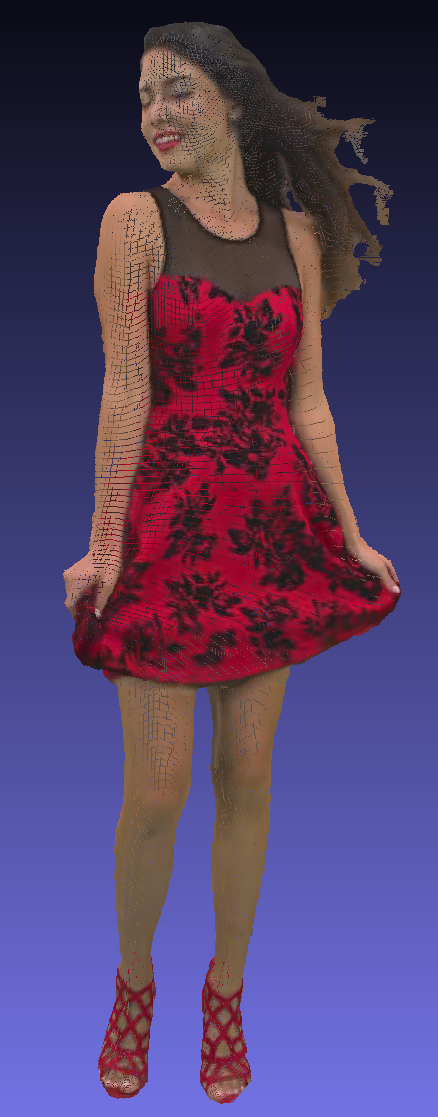}}
\hspace{0.7cm}
\subfloat[B+E]{%
    \includegraphics[width=2.3cm,height=4.2cm]{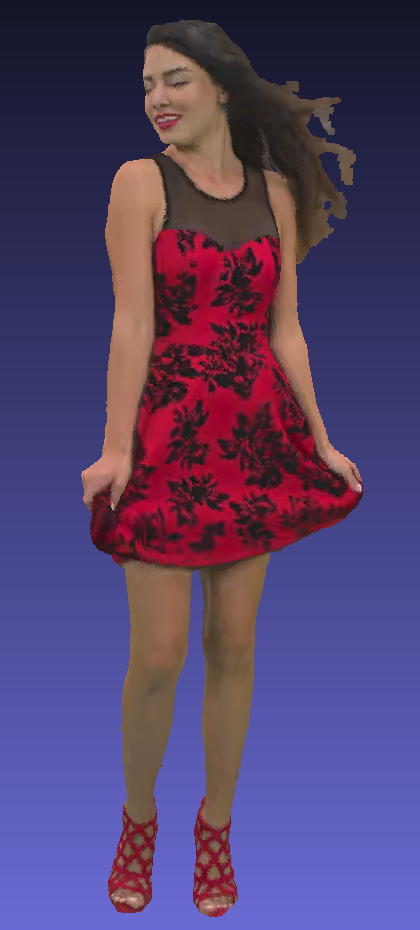}}
\caption{Visual comparison of Base-only and Base-plus-Enhanced (B+E) reconstructions.}
\Description{Side-by-side comparison of the Base-only point-cloud reconstruction and the Base-plus-Enhanced reconstruction obtained by stitching the decoded complementary point sets.}
\label{fig:visual_quality}
\vspace{-1em}
\end{figure}

\begin{table}[t]
\centering
\caption{Visual quality for \textit{Red and Black}.}
\label{tab:visual_quality}
\small
\setlength{\tabcolsep}{3pt}
\renewcommand{\arraystretch}{1.15}
\begin{tabularx}{\columnwidth}{@{}l>{\centering\arraybackslash}X>{\centering\arraybackslash}X@{}}
\toprule
\textbf{Configuration} & \textbf{PSNR (dB)} & \textbf{SSIM} \\
\midrule
Base only & 18.12 & 0.6055 \\
Base + Enhanced & \textbf{22.45} & \textbf{0.7632} \\
Absolute gain & +4.33 & +0.1577 \\
\bottomrule
\end{tabularx}
\vspace{-2em}
\end{table}

\begin{figure*}[t]
    \centering

    \begin{subfigure}[t]{\textwidth}
        \centering
        \includegraphics[width=0.9\textwidth]{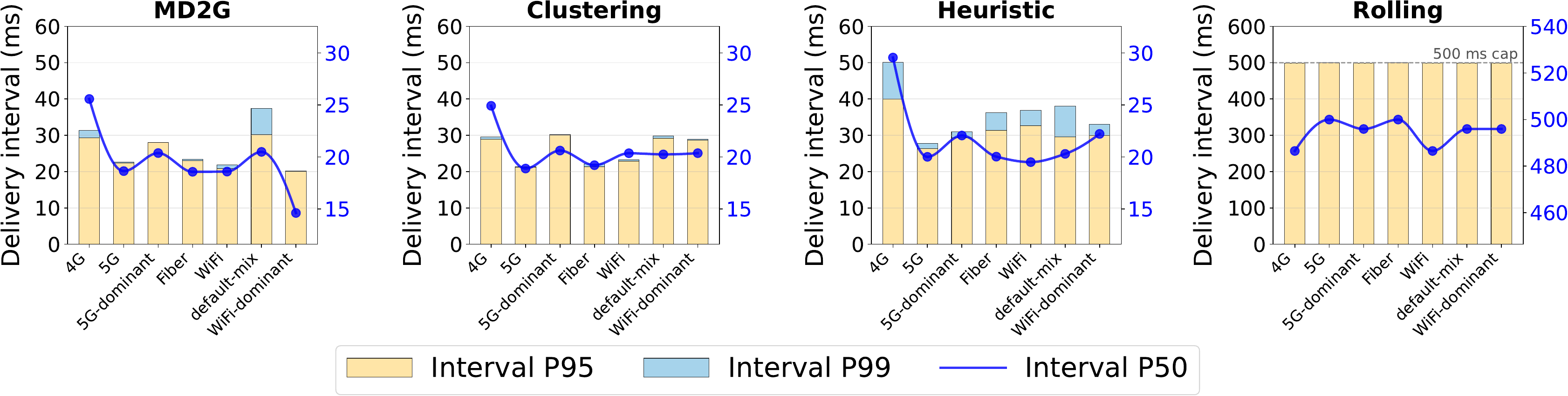}
        \caption{20 users}
        \label{fig:ttfb_20}
    \end{subfigure}
\vspace{-1.5em}
    \begin{subfigure}[t]{\textwidth}
        \centering
        \includegraphics[width=0.9\textwidth]{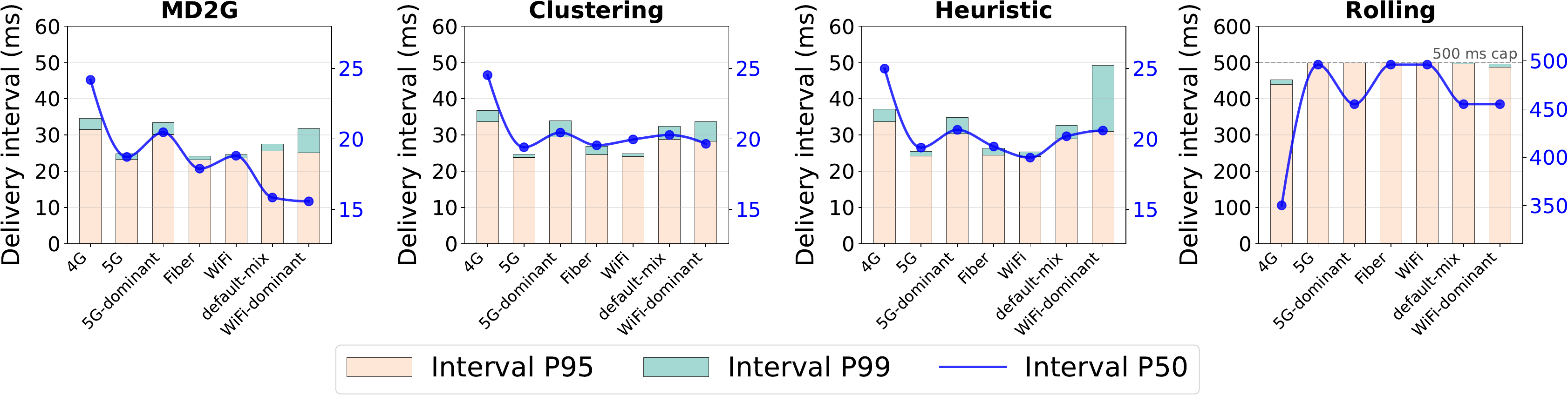}
        \caption{100 users}
        \label{fig:ttfb_100}
    \end{subfigure}
    \caption{Receiver-side delivery interval at 20 and 100 users. The lower bar segment ends at $P_{95}$, the full bar height marks $P_{99}$, and the blue curve shows $P_{50}$. The dashed line marks the 500\,ms reporting cap.}
    \Description{Delivery interval percentiles for four strategies across seven access profiles at 20 and 100 users.}
    \label{fig:ttfb_latency}
    \vspace{-1em}
\end{figure*}

\textbf{Heuristic} is a same-substrate rule-based baseline inspired by joint grouping and resource-allocation designs~\cite{zhang2021joint}. It groups users into channel-quality bins and assigns versions and bandwidth budgets at group level.

\textbf{Clustering} is a same-substrate FoV-driven baseline inspired by prediction-aware grouping~\cite{perfecto2020taming}. It clusters users by viewing similarity to increase reusable demand.

\textbf{Rolling} follows a client-local adaptive delivery design over HTTP/DASH~\cite{li2023toward}. Each client independently adapts and fetches its media under changing bandwidth and FoV conditions, with no application-layer sharing across receivers.

\textbf{Model efficiency.} Table~\ref{tab:model_profile} profiles Teacher-512 and Student-128 on a single CPU thread. Student-128 uses about one seventh of the parameters and memory of Teacher-512, while inference latency falls from 0.083--0.084\,ms to 0.039--0.041\,ms and inference rate approximately doubles. Both models execute far below the one-second control interval, leaving substantial compute headroom for relay processing.

\textbf{Visual reconstruction quality.} Figure~\ref{fig:visual_quality} compares the Base-only reconstruction with the Base-plus-Enhanced reconstruction obtained by stitching the decoded complementary point sets. Base and Enhanced remain separate atlas streams during delivery; receiving both adds their delivery rates, while reconstruction quality is obtained by stitching the decoded point sets rather than by forming a single combined bitstream. Table~\ref{tab:visual_quality} quantifies the improvement: PSNR increases from 18.12 to 22.45\,dB, a gain of 4.33\,dB, while SSIM increases from 0.6055 to 0.7632.

\subsection{Performance in Various Network Environments}
\label{subsec:perf_all}
We evaluate 10, 20, 40, 60, and 100 concurrent users under homogeneous and heterogeneous access. Each run lasts 120 seconds, and every plotted point aggregates three trials under the same trace, topology, and measurement procedure.

All users consume the \textit{Red and Black} sequence, partitioned into complementary Base and Enhanced point sets and encoded as separate V-PCC atlas streams. Relay decisions are updated every 1.0\,s using current FoV overlap, receiver capability, and bandwidth adequacy derived from passively observed delivery rates. The bandwidth signal is smoothed with an exponential moving average before entering the controller.

\textbf{Steady-state delivery interval.} Figure~\ref{fig:ttfb_latency} reports the receiver-side inter-arrival interval of useful media objects after the 30\,s warm-up. At both 20 and 100 users, the three MoQ strategies remain well below the HTTP/DASH independent-delivery reference. MD2G-Cast keeps $P_{99}$ below 40\,ms across all seven evaluated access profiles. Clustering shows a comparable tail, whereas Heuristic reaches approximately 50\,ms in its most demanding setting. At 20 users, Rolling reaches the 500\,ms reporting cap across the evaluated profiles. At 100 users, its $P_{99}$ remains roughly 450--500\,ms, with most profiles at or near the cap. Its occasionally lower median therefore does not translate into stable tail behavior.

The Rolling comparison reflects both its independent-delivery model and its HTTP/DASH transport. It must not be interpreted as a transport-only result. For isolating relay-control effects, the same-substrate MoQ comparison is the relevant comparison: MD2G-Cast maintains a controlled delivery tail while coordinating grouping and Enhanced admission at the relay. We therefore interpret Fig.~\ref{fig:ttfb_latency} jointly with the enhancement-exposure, aggregate-link-load, and system-utility results.

\begin{figure}[t]
    \centering
    \includegraphics[width=1.0\linewidth]{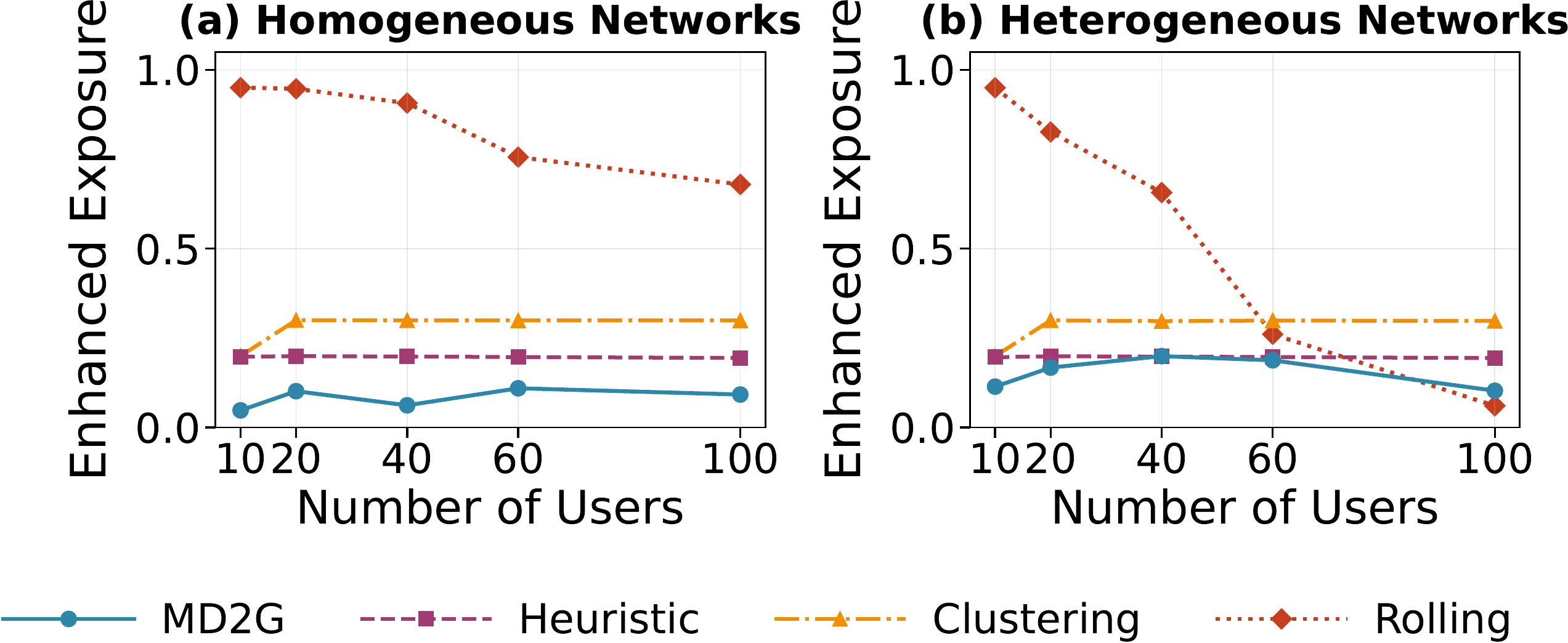}
    \caption{Enhanced exposure across user scales under homogeneous and heterogeneous access.}
    \Description{Enhanced exposure across five user scales under homogeneous and heterogeneous access.}
    \label{fig:enhanced_ratio}
    \vspace{-1em}
\end{figure}

\textbf{Enhanced delivery exposure.} Figure~\ref{fig:enhanced_ratio} reports the fraction of expected-client steady-state user-time for which Enhanced delivery is active. Rolling exposes Enhanced delivery for approximately 0.95 of user-time at low load. Its exposure remains high under homogeneous access but falls sharply under heterogeneous access, from about 0.95 at 10 users to about 0.06 at 100 users. Heuristic remains near 0.20, while Clustering stays around 0.30 from 20 users onward.

MD2G-Cast uses Enhanced delivery more selectively. Under homogeneous access, its exposure remains approximately 0.05--0.11. Under heterogeneous access, it rises to roughly 0.10--0.20 between 10 and 60 users before returning to about 0.10 at 100 users. Enhanced exposure alone does not establish delivered quality because an admitted object is useful only if it arrives in time for playback. The combination of selective exposure, controlled delivery intervals, and the utility results therefore supports adaptive admission rather than maximizing Enhanced activity.

\begin{figure}[t]
    \centering
    \includegraphics[width=1.0\linewidth]{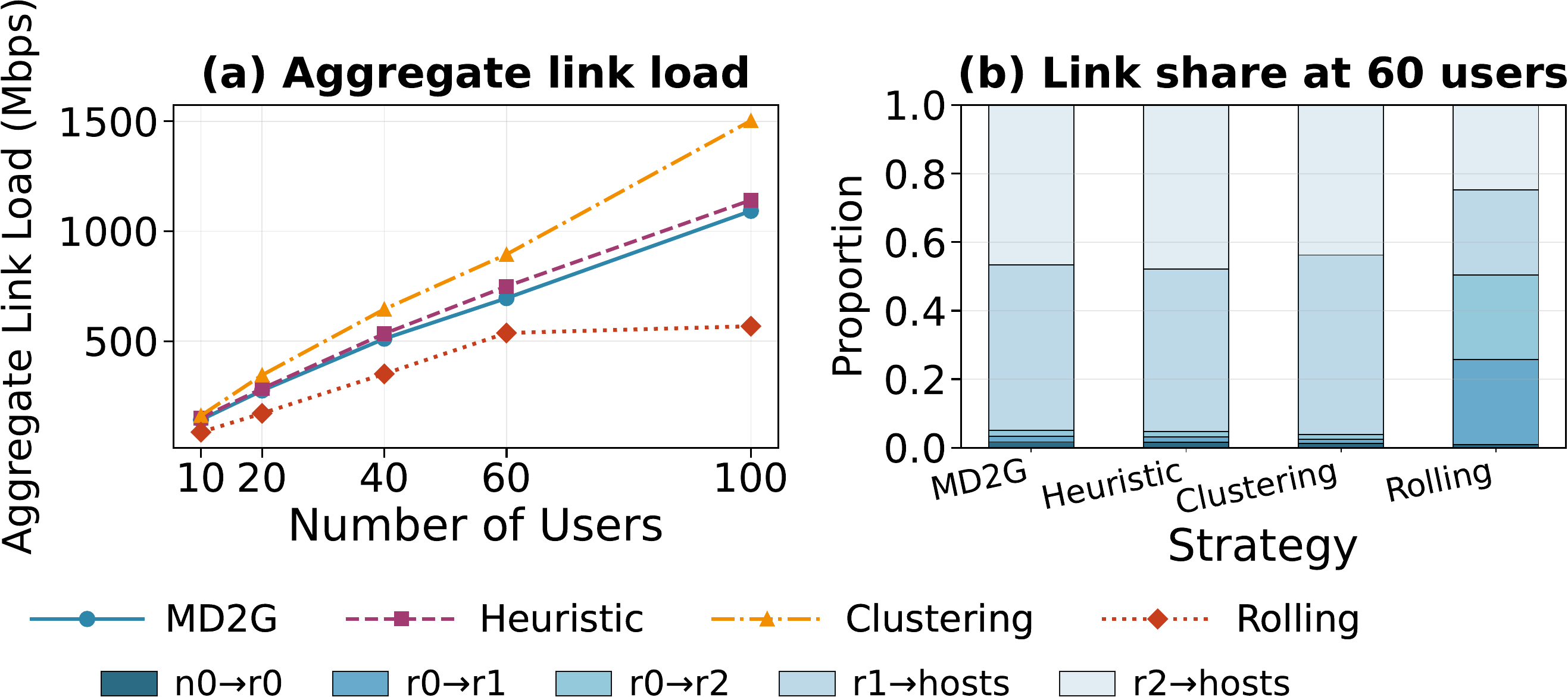}
    \caption{Aggregate link load across user scales and normalized link shares at 60 users.}
    \Description{Aggregate link load across user scales and normalized traffic shares across five monitored links at 60 users.}
    \label{fig:throughput_total}
    \vspace{-1em}
\end{figure}

\textbf{Aggregate link load.} Figure~\ref{fig:throughput_total}(a) reports the carried traffic summed over the five monitored directed links. This quantity represents aggregate network work rather than application throughput or utility. At 100 users, Clustering incurs approximately 1.50\,Gbps, compared with about 1.10\,Gbps for MD2G-Cast and 1.14\,Gbps for Heuristic. MD2G-Cast therefore reduces aggregate link load by about 27\% relative to Clustering at the largest evaluated scale.

Rolling reaches only about 0.56\,Gbps at 100 users, but this lower value does not indicate greater efficiency. Its carried traffic plateaus under load, consistent with congestion-limited delivery, a near-cap delivery interval, and very low system utility.

Figure~\ref{fig:throughput_total}(b) shows link placement at 60 users. Rolling concentrates most network work on downstream branches because each client retains an independent delivery path. The MoQ strategies share upstream delivery and fan out at the leaf relays, shifting traffic toward the edge. Aggregate link load should therefore be read together with delivery continuity and system utility.

\begin{figure}[t]
    \centering
    \includegraphics[width=1.0\linewidth]{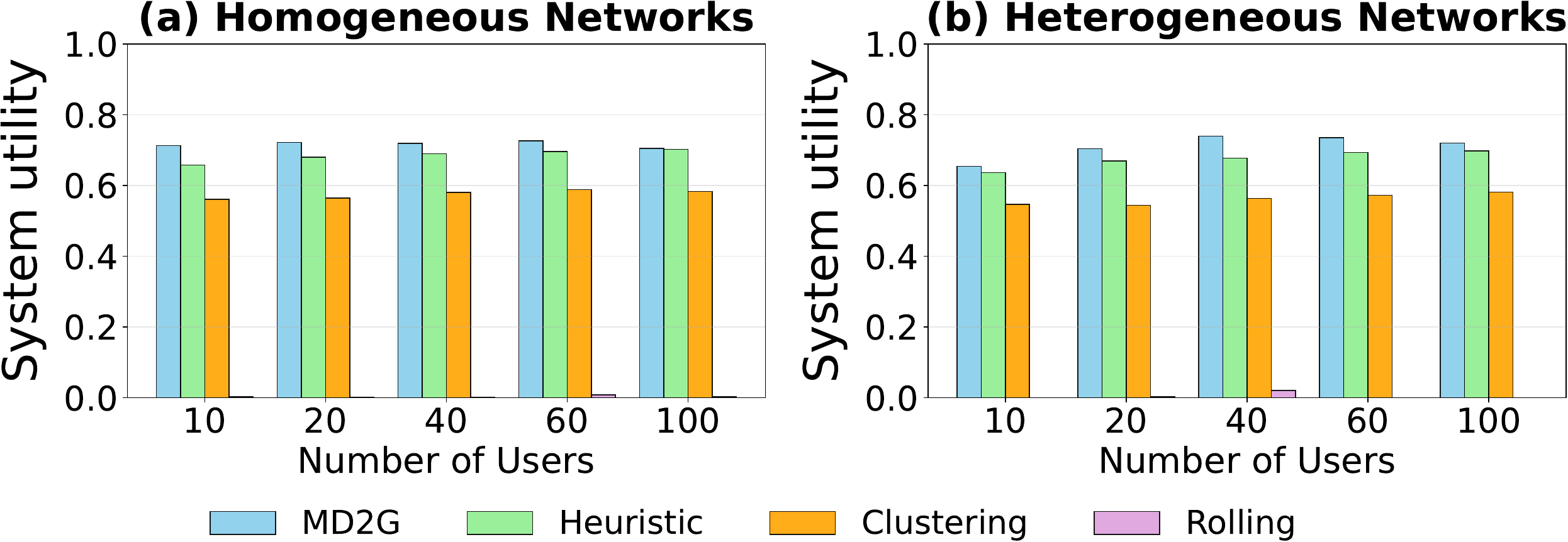}
    \caption{Mean system utility across homogeneous and heterogeneous access profiles after the 30\,s warm-up.}
    \Description{Mean system utility for four strategies across five user scales under homogeneous and heterogeneous access.}
    \label{fig:qoe_bar}
\vspace{-1em}
\end{figure}

\textbf{System utility.} Figure~\ref{fig:qoe_bar} reports the expected-client mean of the clipped utility in Eq.~\eqref{eq:system_utility}. At each user scale, the homogeneous panel aggregates 4G, 5G, fiber, and Wi-Fi, while the heterogeneous panel aggregates 5G-dominant, default-mix, and Wi-Fi-dominant access. Under homogeneous access, MD2G-Cast achieves the highest or tied-highest mean utility at every evaluated scale. Its advantage is most visible at moderate loads, while MD2G-Cast and Heuristic become nearly tied at 100 users. Clustering remains consistently lower, and Rolling contributes little utility because aggressive Enhanced exposure does not compensate for its poor delivery continuity.

Under heterogeneous access, MD2G-Cast attains the highest mean utility at every evaluated user scale. The separation from Heuristic grows from light load toward the intermediate scales and narrows again at 100 users, whereas the gap to Clustering remains clearer. The non-monotonic trend is consistent with two competing effects of scale: a larger population creates more opportunities for reusable demand, but increasing concurrency also raises contention for shared and access-link capacity. MD2G-Cast therefore does not obtain its utility by maximizing any single component. It benefits from reuse when sharing is available and becomes more selective when additional refinement would impose excessive network cost. We restrict this interpretation to the workloads and access profiles evaluated here.

\subsection{Matched Relay Control Ablation}
\label{subsec:ablation}

We isolate the relay control mechanism from the common MoQ substrate under the default heterogeneous mix with 60 users. All variants use the same media, topology, access traces, 120\,s duration, 30\,s warm-up, measurement method, and three matched trials. Only the relay decision mechanism changes. Figure~\ref{fig:qoe_bar} aggregates several heterogeneous access profiles, whereas this ablation evaluates default heterogeneous mix alone, so the corresponding utility values need not coincide.

\begin{table}[t]
\centering
\caption{Matched relay-control ablation under the default heterogeneous mix with 60 users.}
\label{tab:policy_ablation}
\small
\setlength{\tabcolsep}{2pt}
\renewcommand{\arraystretch}{1.08}
\begin{tabular*}{\columnwidth}{@{\extracolsep{\fill}}lcccc@{}}
\toprule
\textbf{Variant} &
\textbf{Utility} &
\(\boldsymbol{\Delta U}\) &
\(\boldsymbol{R_q}\) &
\(\boldsymbol{R_b}\) \\
\midrule
Full MD2G-Cast       & $0.716\pm0.004$ & ---      & 0.777 & 0.005 \\
Fixed grouping       & $0.712\pm0.003$ & $-0.004$ & 0.770 & 0.002 \\
Random feasible action & $0.703\pm0.014$ & $-0.013$ & 0.773 & 0.075 \\
Rule-only controller & $0.718\pm0.006$ & $+0.002$ & 0.782 & 0.007 \\
\bottomrule
\end{tabular*}
\vspace{-1em}
\end{table}

The mean grouping-efficiency term is $R_o=1.0$ for all variants in this setting and is therefore omitted from Table~\ref{tab:policy_ablation}. Fixed grouping changes mean utility only slightly. Random feasible action retains delivered quality close to Full, with $R_q=0.773$ versus $0.777$, but raises bandwidth inefficiency from $R_b=0.005$ to $0.075$. Its utility loss therefore arises primarily from bandwidth-inefficient choices rather than a large reduction in delivered quality.

The deterministic rule controller remains comparable to Full MD2G-Cast and is slightly higher in mean on this workload. The ablation therefore supports structured relay control but does not establish PPO superiority over deterministic control.

\textbf{Cross-metric interpretation.} Figures~\ref{fig:ttfb_latency}--\ref{fig:qoe_bar} show why no single metric characterizes the system. Rolling keeps Enhanced delivery active for most user-time at light load, yet its delivery interval remains close to the reporting cap, so high exposure does not imply useful refinement. Clustering carries the largest aggregate link load at scale without attaining the highest utility. MD2G-Cast instead combines selective Enhanced admission with controlled delivery intervals and lower network work than Clustering, producing the highest mean heterogeneous utility and the highest or tied-highest homogeneous utility.
\section{Related Work}
\label{sec:related_work}

\textbf{Per-receiver volumetric adaptation.}
GROOT provides a real-time pipeline for high-fidelity volumetric streaming, while ViVo exploits visibility to reduce mobile delivery cost \cite{lee2020groot,han2020vivo}. Vues uses multiview transcoding to balance bandwidth, rendering cost, and viewpoint flexibility, and YuZu combines neural enhancement with network and compute adaptation \cite{liu2022vues,zhang2022yuzu}. MetaStream extends this line of work to live capture, creation, delivery, and rendering \cite{guan2023metastream}. These systems improve the efficiency or quality of an individual delivery session. MD2G-Cast addresses a different bottleneck. It coordinates content reuse across concurrent receivers at the relay.

\textbf{Multi-user sharing and multicast.}
Prior multicast systems jointly optimize receiver grouping, version selection, and bandwidth allocation for conventional or 360-degree video \cite{perfecto2020taming,zhang2021joint, mahmoud2023survey}. Volumetric systems have subsequently explored cross-layer sharing, soft multicast, viewpoint-based clustering, and hybridized delivery~\cite{zhang2022m5,ueno2023soft,lai2024socially,liu2024muv2}. These studies establish that cross-user reuse can reduce repeated transmission. Many of these designs are tailored to a particular wireless channel, representation, or grouping rule. MD2G-Cast instead exposes volumetric and receiver state to a relay controller that jointly selects grouping and optional refinement admission.

\textbf{Complementary content and relay execution.}
Complementary and layered representations support fine-grained adaptation by making additional reconstruction detail optional when resources permit. Swift demonstrates this principle for learned layered video coding, while knowledge distillation provides a general mechanism for reducing online model cost~\cite{dasari2022swift,gou2021knowledge}. MD2G-Cast applies a related systems principle to shared Base-content fan-out and receiver-specific Enhanced admission at MoQ relays, with Base and Enhanced encoded as complementary point-set streams.

\textbf{Media over QUIC.}
MoQ provides object-oriented publish and subscribe delivery
over QUIC and supports relay-assisted distribution~\cite{10.1145/3587819.3593937,gurel2024media}. Recent work has also studied object prioritization within MoQ~\cite{gurel2024priority}. These efforts define and evaluate the transport substrate. They do not determine how volumetric receivers should be grouped or how receiver capability, FoV overlap, and bandwidth headroom should control grouping and Enhanced admission. MD2G-Cast contributes this application-aware relay decision layer without changing the underlying MoQ or QUIC wire behavior.
\section{Discussion and Conclusion}
\label{sec:Conclusion}

\textbf{Interpretation.} The controlled MoQ comparisons separate relay decision logic from the benefit of multicast delivery itself. Heuristic and Clustering use the same MoQ substrate, Base/Enhanced media representation, topology, traces, and measurement procedure as MD2G-Cast, so their differences primarily reflect how grouping and Enhanced admission are controlled. The matched ablation provides a second view of this separation. Fixed grouping produces a small loss, random action selection is less effective and less stable, and the deterministic rule controller remains comparable to PPO. The evidence therefore supports structured relay coordination as the central contribution while showing that the particular optimizer can vary with the workload.

\textbf{Design implication.} MD2G-Cast deliberately separates fixed media preparation from adaptive relay control. Base and Enhanced objects are created once before publication, while runtime adaptation changes only multicast grouping and Enhanced admission. This separation keeps media construction and feasibility outside the optimizer, allows learned and deterministic controllers to share the same relay interface, and lets the relay exploit cross-user overlap that independent clients cannot coordinate on their own.

\textbf{Scope and deployment.}
MD2G-Cast relies on reusable FoV structure and offers fewer sharing opportunities when receivers request unrelated content. Its system utility is an objective proxy combining grouping efficiency, delivery-derived quality, and bandwidth inefficiency. The quality term incorporates delivery-interval and stall penalties, but the resulting objective does not replace subjective quality assessment. The current prototype uses one V-PCC sequence and a two-leaf relay hierarchy. Additional scenes, subjective studies, and coordination across deeper relay trees remain future work. MD2G-Cast operates above the MoQ transport interface and does not modify QUIC or MoQ wire behavior.

\textbf{Conclusion.} MD2G-Cast introduces an application-aware relay coordination layer for scalable volumetric delivery over MoQ. Rather than treating multicast as a fixed transport choice, it makes cross-user reuse and optional refinement explicit relay decisions. At the 20- and 100-user loads reported in Fig.~\ref{fig:ttfb_latency}, MD2G-Cast keeps the receiver-side $P_{99}$ delivery interval below 40\,ms across all seven access profiles; at 100 users, it reduces aggregate link load by about 27\% relative to Clustering. It achieves the highest or tied-highest mean utility across homogeneous user scales and the highest mean utility across heterogeneous user scales. Together, these results support a relay control layer whose policy realization can evolve without changing the encoded Base/Enhanced objects or the underlying MoQ transport.

\clearpage
\bibliographystyle{ACM-Reference-Format}
\balance
\bibliography{Reference}

\end{document}